\documentclass[showpacs,preprintnumbers,amsmath,amssymb]{revtex4}
\usepackage{graphicx}
\usepackage{dcolumn}
\usepackage{bm}

\newcommand{\pdf}{\textsc{PDF}}
\begin{document}


\title{Statistical Equilibria of Uniformly Forced Advection Condensation}
\author
{Jai Sukhatme$^1$
 and Raymond T. Pierrehumbert$^2$\\
 $^1$ Mathematics Department, University of Wisconsin-Madison, Madison, WI 53706 \\
 $^2$ Department of Geophysical Sciences,
  University of Chicago, Chicago,  IL 60637 }

\date{\today}

\begin{abstract}
We examine the state of statistical equilibrium attained by a uniformly
forced condensable substance subjected to advection in a periodic
domain. In particular, we examine the 
probability density function (\pdf{}) of the condensable substance
in the limit of rapid condensation. The constraints imposed by this limit are pointed out and are shown to result in
a \pdf{} --- whenever the advecting velocity field admits a diffusive representation --- that features a
peak at small values,
decays exponentially and terminates in a
rapid "roll-off" near saturation. Possible physical implications of this 
feature as compared to a \pdf{} which continues to decay slowly are pointed out. A
set of simple numerical exercises which employ lattice maps for purposes of advection are 
performed to test these features.
Despite the simplicity of the model, the derived \pdf{} is seen to compare
favourably with \pdf{}'s constructed from isentropic specific humidity 
data. 
Further, structure functions associated with the condensable field are seen to scale anomalously with 
near saturation of the scaling exponents for high moments
--- a feature which agrees with studies of high resolution aircraft data. 
\end{abstract}

\maketitle

\section{Introduction}

{\it Advection-Diffusion-Condensation} (ADC) is a variant of the familiar passive advection-diffusion problem
\cite{Falk}, wherein apart from molecular diffusion a tracer is subject to an additional sink associated 
with the process of condensation. 
A study of the interplay of these processes is motivated by the need to understand the large 
scale distribution of water vapor in the troposphere \cite{Sherwood},\cite{SH},\cite{Ray-grl}. 
Indeed, the prominent role played by
water vapor in various problems related the Earth's climate \cite{Ray-Nature} --- for example, via the water 
vapor feedback 
in the greenhouse effect \cite{HS} --- makes this an important issue. 
In this context, ADC serves as an idealized problem for 
water vapor in the troposphere \cite{Ray-2}.
The mixing ratio $q(\vec{x},t)$ of a scalar tracer subject to advection, diffusion and condensation is
governed by

\begin{equation}
\frac{\partial q}{\partial t} + (\vec{u} \cdot \nabla)q = \eta \nabla^2 q + S(q,q_s) + F
\label{1}
\end{equation}
here $\vec{u}$ and $F$ are the advecting velocity field 
and the forcing respectively and $\eta$ is the diffusivity of the condensable substance.
In the present work we shall assume the domain to be spatially bi-periodic and restrict our attention to the
nondiffusive limit, $\eta=0$.
The problem thus consists of the advection of a passive tracer supplied by the source $F$, and removed by a sink
associated with condensation.
This sink 
is represented as

\begin{eqnarray}
S = -\frac{1}{\tau}[q - q_s(\vec{x})]  \quad \textrm{if} \quad q>q_s \nonumber \\
= 0 \quad \textrm{if} \quad q \le q_s
\label{2}
\end{eqnarray}
where $q_s(\vec{x})$ is the saturation mixing ratio --- a prescribed function. 
Hence with $\eta=0$, the ADC problem is closely related to the interaction of smooth advection with linear damping \cite{Chertkov1}. 
The ADC problem also bears a similarity recent studies of "active" processes when coupled with 
fluid advection, such as 
the evolution of chemically \cite{Neufeld} or 
biologically \cite{Abr} 
active substances and the issue of phase separation in immiscible fluids \cite{Bert}. \\ 

Our aim is to examine the
state of statistical equilibrium attained by (\ref{1}) and (\ref{2})
in the limit $\tau \rightarrow 0$ or that of {\it rapid condensation}. 
This limit implies that
upon advection if $q(\vec{x},t) > q_s(\vec{x})$, then the parcel's mixing ratio is instantaneously reset
to the saturation mixing ratio at that location. Of course, two immediately
apparent opposing long time limits of
(\ref{1}) and (\ref{2}) are : (i) Uniform forcing with $u=0 
\Rightarrow q(\vec{x},t) \rightarrow q_s(\vec{x})$ and (ii) A sufficiently mixing flow with
$F=0 \Rightarrow q(\vec{x},t) \rightarrow \min_{\forall \vec{x}} [q_s(\vec{x})]$. Hence, to
achieve a non-trivial state of statistical equilibrium we require both $\vec{u},F \neq 0$. \\

In the absence of diabatic effects, parcel motion in the midlatitude troposphere 
is restricted to two-dimensional surfaces of constant potential temperature, i.e. to 
isentropic surfaces \cite{Hoskins}.
Our objective is to understand the probability density function (\pdf{}) of the water vapor mixing ratio
along these midlatitude isentropic surfaces \cite{Ray-4}. 
On average the thermal structure of the midlatitude troposphere is such that as one
progresses polewards, the temperature
along isentropic surfaces decreases in a fairly linear manner. 
Given the sensitivity of the saturation mixing ratio (via the Clausius-Clapeyron  
relation) to the temperature, following \cite{Ray-2}
we take $q_s$ to vary exponentially with $y$. As we are in a periodic 
domain,
$q_s(\vec{x})=q_s(y)=\exp(-\alpha |y-\frac{L}{2}|)$ ($0 \le y \le L$) where $\alpha > 0$ and 
increasing $\alpha$ yields progressively steeper profiles which fall 
off symetrically from $y=\frac{L}{2}$. This yields an idealized model problem which retains
the essence of the much more complex atmospheric problem that motivates our work. Even though
the real atmosphere does not conform exactly to the idealizations, progress can be made through
a detailed solution of this model problem.\\

\section{The PDF in the limit of rapid condensation}

\subsection{General formulation and boundary conditions} 

Before we proceed to the \pdf{} in the general case,
consider the limit alluded to in the Introduction. Specifically,
$\vec{u}=0$ with constant forcing. On one hand this case is quite straightforward but on the other it
exhibits some pathologies inherent in dealing with the \pdf{}. From (\ref{1}) at a location $x_0,y_0$ for $t>
t_1=q_s(x_0,y_0)/F$ we have

\begin{equation}
q(x_0,y_0,t) = q_s(x_0,y_0) + F \tau [ 1 - \exp(\frac{-(t-t_1)}{\tau}) ] 
\label{e1}
\end{equation}
As (\ref{e1}) shows, in general the domain will become supersaturated. But for rapid condensation, as claimed,
$q(x_0,y_0,t) \rightarrow q_s(x_0,y_0) ~\forall (x_0,y_0)$. Note that the limit
of saturation is approached from above. Physically this implies that as $\tau \rightarrow 0$, we simultaneously have 
$q(x,y,t) \rightarrow q_s(x,y)$ such that the condensation sink balances the forcing. \\

To estimate the \pdf{}
(denoted by $P=P(x,y,q,t)$), we decompose the stationary
solution as $P(x,y,q)=P_1(x,y) F_1(q|x,y)$ where $F_1(q|x,y)$ is the conditional \pdf{} 
of $q$ conditioned on $(x,y)$. From the preceeding discussion, $F_1(q|x,y)=\delta[q - q_s(x,y)]$. As $\vec{u}=0$, 
for an unbiased estimate it is natural to take $P_1(x,y)$ to be uniform resulting 
in $P(x,y,q)=\delta[q - q_s(x,y)]$. In essence 
we end up with a $\delta$ function supported on the fixed point of the dynamical system (\ref{1}) with 
$\vec{u}=0$ \cite{Cycle}. In this particular case, the system does not have any other invariant \pdf{} and
the $\delta$ function \pdf{} is meaningful (as will be seen later), but in general to avoid such complications we restrict
our attention to functions that are reasonably smooth. \\

Proceeding to the general case 
\footnote{We are grateful to the referee who pointed out problems in our original formulation of the
\pdf{} equation. Indeed, this section would be much weaker if not for his comments.}, the 
Liouville or transport equation satisfied by the \pdf{} is

\begin{eqnarray}
\frac{\partial P(x,y,q,t)}{\partial t} + \frac{\partial (u_iP)}{\partial x_i} +
\frac{\partial[(F+S)P]}{\partial q} = 0 \nonumber \\
\textrm{with}~\int_{\vec{x}} {\int_{0}}^{\infty} P(x,y,q,t) ~d\vec{x}~dq = 1 ~ \forall t
\label{3a}
\end{eqnarray}
as in general, i.e. outside the limit of rapid condensation, $0 \le q(x,y,t) \le \infty$ and apart from periodicity in $\vec{x}$
we require $P(x,y,\infty,t) = 0$. \\

In $(x,y,q)$ space the above can be viewed as resulting from a 
{\it probability current} $\vec{U} P$ where $\vec{U} = [u,v, (F+S) ]$. An immediate consequence  
of rapid condensation is a restriction on $q$ i.e. $q(x,y,t) \le q_s(x,y)$. Till now our choice of forcing has been 
completely arbitrary, at this stage we restrict ourselves to a forcing which is positive definite and 
also satisfies $\frac{\partial F}{\partial q}=0$.
This implies, whatever the initial 
condition on $q$ is, after a finite time $\min{(q)}=\min{(q_s)}$. Hence, $\min{(q_s)} \le q(x,y,t) \le q_s(x,y)$ and
using (\ref{2}), (\ref{3a}) reduces to

\begin{eqnarray}
\frac{\partial P(x,y,q,t)}{\partial t} + \frac{\partial (u_iP)}{\partial x_i} +
\frac{\partial(FP)}{\partial q} = 0 \nonumber \\
\textrm{with}~\int_{\vec{x}} {\int_{\min{(q_s)}}}^{q_s(x,y)} P(x,y,q,t) ~d\vec{x}~dq = 1 ~ \forall t
\label{3b}
\end{eqnarray}
and, as noted, apart from the restriction on the forcing function, we require $P$ to be relatively smooth. 
Also, note that now the domain is periodic in $(x,y)$ and is bounded by the surfaces $q=\min{(q_s)}$ and
$q = q_s(x,y)$. Further, periodicity and the imposition of rapid condensation imply $[\vec{U}P]|_{\partial \bf{D}} = 0$
and in essence (\ref{3b}) represents the evolution of a \pdf{} via an incompressible flow within a bounded
domain with the normalization constraint implicity providing the required boundary condition.
Interestingly, similar integral
formulations for the
PDF arise in studies of spiking neurons (see for example Fusi \& Mattia \cite{Fusi} and the references
therein). \\

\subsection{Solutions for present saturation profile}

In the present case, as the saturation mixing ratio is purely a function of $y$, if the forcing is also taken to be 
of the form $F=F(y)$ (in fact we will focus on the uniformly forced case) ---
it is reasonable to look for solutions which
are independent of $x$. Of course, it is the realization of statistical equilibrium
without the presence of
gradient fields (as $\eta =0$) that makes
progress possible in the present case \footnote{A similar absence of gradient 
fields in equilibrium has been indirectly exploited in
the study of advection with linear damping \cite{Chertkov1}.}. This circumvents the need 
to estimate conditional expectations
of the scalar dissipation (or diffusion) which complicate advection-diffusion problems \cite{P},\cite{Dopazo-rev},
\cite{jai1}. \\ 

\subsubsection{The uniform case}

Inspecting (\ref{3b}) it is evident that without any further assumptions, as per the usual situation involving 
the evolution of a \pdf{} via an incompressible flow, a valid stationary solution is that of a uniform distribution in $(y,q)$. 
To obtain the $\pdf{}$ of $q$ from this uniform distribution we estimate
the probability that $q<Q$ as a function of $Q$
(i.e. $Q$ represents the sample space variable corresponding
to $q$).
Utilizing the rapid condensation normalization constraint,
$P(q,y)$ has to be integrated with respect to the
saturation curve $q_s(y)$. Therefore

\begin{equation}
Pr(q<Q) = \frac{I_1+I_2}{I_t}
\label{7a}
\end{equation}
Here $I_1,I_2$ correspond to the regions
$y \le Y$ and $y > Y$ where $Y=Z(Q)$ ($Z$ being the inverse of $q_s(y)$). In the present case, as $q_s$ is symmetric
about $y=\frac{L}{2}$ we need only consider
half of the domain ($L/2 \le y \le L$)
with $q_s=\exp(-\alpha y)$, i.e. $Z(Q) = -\log(Q)/\alpha$. Specifically,

\begin{eqnarray}
I_1 = {\int_{L/2}}^{Z(Q)} {\int_{Q_2}}^{Q}~ P(q,y) ~dqdy \nonumber \\
I_2 = {\int_{Z(Q)}}^{L} {\int_{Q_2}}^{q_s(y)}~ P(q,y) ~dqdy
\label{7b}
\end{eqnarray}
where $Q_2=\min(q_s)$. $I_t$ in (\ref{7a}) is the same as $I_2$ but with $L/2$ as the lower limit of integration in the outer
integral, which ofcourse is nothing but the normalization constraint in (\ref{3b}). 
Hence,

\begin{equation}
\pdf{}(Q) = \frac{d (I_1+I_2)}{dQ}  
\label{11aa}
\end{equation}
Substituting $P(y,q)=$ const. in (\ref{7b}) and (\ref{11aa}) yields $\pdf{}(Q) \sim Z(Q)$ i.e. 
$\pdf{}(Q) \sim \log(Q^{\frac{-1}{\alpha}})$ which can be seen in the upper panel of Fig. (\ref{fignew1}). 
In essence, for a uniform distribution the constraint of rapid condensation forces the \pdf{} to reflect the 
saturation profile. \\

\subsubsection{The eddy-diffusion case}

We now assume that the the effect of the fluctuating velocity in the Liouville equation admits a diffusive representation
 --- say by an eddy diffusivity $\kappa_e$.  This would be the case if the parcel trajectories consisted of
independent Brownian motion, for example. 
Admittedly, this is a fairly severe assumption in that mixing by multiple scale 
velocity fields rarely follows a simple diffusive prescription \cite{MK}. But from 
a tropospheric viewpoint, it is known that the meridional ($y$ - direction) Lagrangian eddy-velocity
correlation function decays quite rapidly (on the order of a few days) \cite{jai3} --- hence, in this
context the assumption
of an eddy diffusivity may not completely unreasonable.  Note that the eddy diffusivity  we refer to here is
an eddy diffusivity applied to probability evolution in $(y,q,t)$ space.  This is not the same as characterizing
mixing by an eddy diffusivity to an evolution equation for a coarse-grained $q$ in $(y,t)$ space.  Indeed,
in \cite{Ray-2} it is shown that the Brownian model yields different coarse-grained $q$ statistics than the
mean-field diffusivity model; it is suggested further that the Brownian motion model constitutes a minimal model
for investigation of the interplay of transport processes with a nonlinear sink term such as condensation. In that
sense, the Brownian model is worth of study in and of itself. \\

\begin{figure}
\includegraphics[width=7.5cm,height=9.5cm]{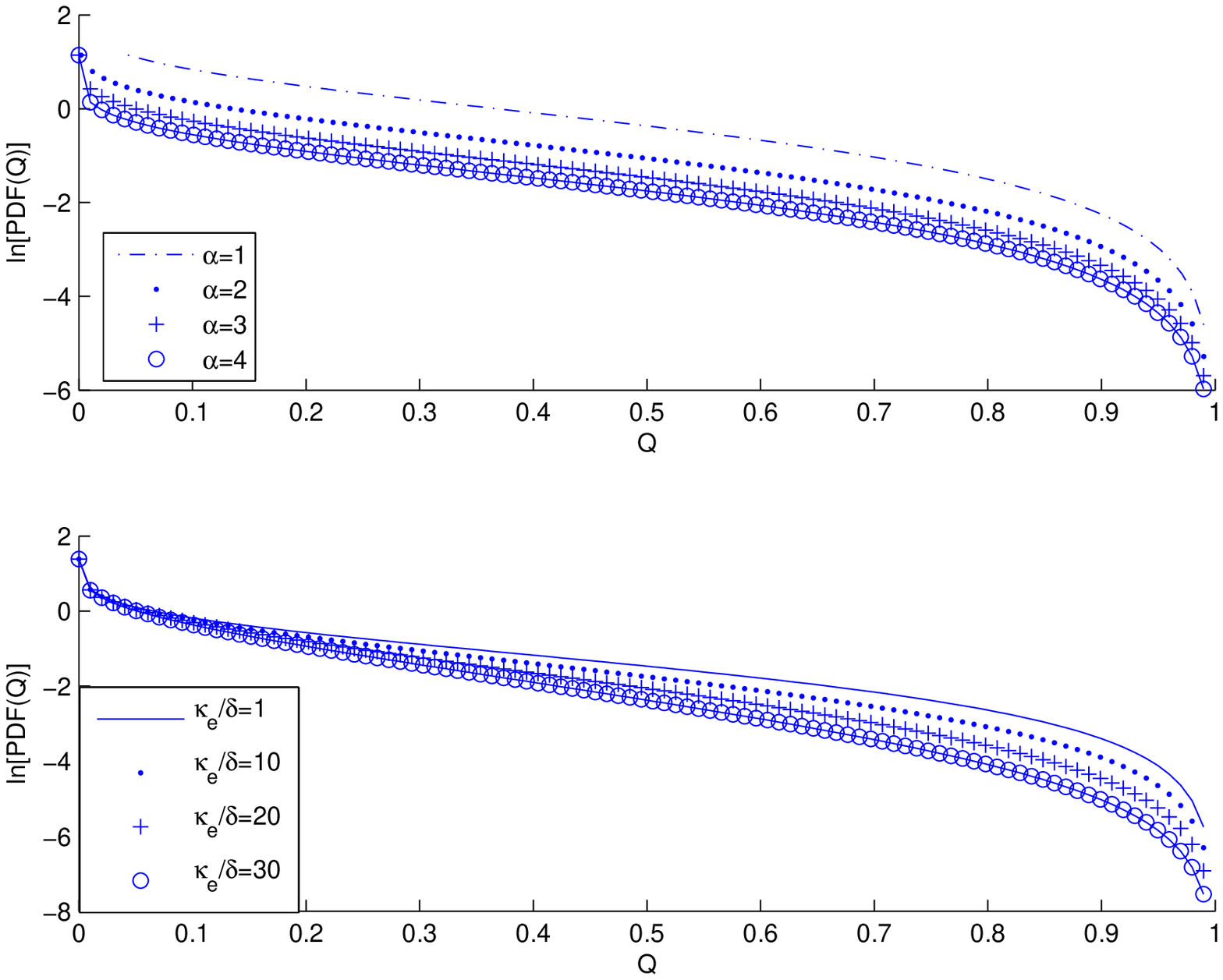}
\caption{\label{fignew1} Upper Panel : The \pdf{} for the uniform case showing the dependence on $\alpha$. Lower Panel : The \pdf{} 
in (\ref{11b}) as a function of $\frac{\kappa_e}{\delta}$ with $\alpha=3$.}
\end{figure}

With the eddy-diffusivity (\ref{3b}) yields 

\begin{equation}
\frac{\partial P}{\partial t} -\kappa_e \frac{\partial^2 P}{\partial y^2} + F(y) \frac{\partial P}{\partial q} = 0
\label{5}
\end{equation}
For uniform forcing $F(y)=\delta$ (a constant), as $P > 0$ and is periodic in $y$, by inspection the ansatz 


\begin{equation}
P(q,y) \sim \exp(-k^2 q)[ \cos(\lambda y)+  \sin(\lambda y)]  ~;~ \textrm{where} ~\lambda=\frac{1}{4},k^2=\lambda ^2 \frac{\kappa_e}{\delta}
\label{7}
\end{equation}
satisfies a stationary form of (\ref{5}). Indeed, for a quantitative estimate we would need to fix the constants that 
arise in (\ref{7}) via the normalization constraint. 
But for a qualititative estimate substituting $P(q,y)$ in (\ref{7b}) and setting $dG/dy=[ \cos(\lambda y)+ \sin(\lambda y)]$ for 
notational simplicity, yields

%
%
\begin{equation}
\pdf{}(Q) \sim ~\exp(-k^2Q) ~[G(Z(Q)) - G(0)] 
\label{11}
\end{equation}
Substituting for $G(y)$

\begin{equation}
\pdf{}(Q) \sim \frac{ \exp(-k^2Q) }{\lambda} ~\{ \sin[\lambda Z(Q)] - \cos[\lambda Z(Q)] + 1 \}
\label{11a}
\end{equation}
and finally substituting for $Z(Q)$  

\begin{equation}
\pdf{}(Q) \sim ~ \frac{ \exp(-k^2Q) }{\lambda} ~\{\sin[\log(Q^{\frac{-\lambda}{\alpha}})] -
\cos[\log(Q^{\frac{-\lambda}{\alpha}})] + 1 \}
\label{11b}
\end{equation}
The dependence of the \pdf{} on $\frac{\kappa_e}{\delta}$ is shown in 
the lower panel of 
Fig. (\ref{fignew1}).
In all cases, 
the \pdf{} has a peak for small $q$ 
decreases exponentially for intermediate
values of $q$ (the slope increases with $\frac{\kappa_e}{\delta}$) and then rolls-off as $q \rightarrow \max{(q_s)}$. \\

Comparing this with the uniform case (upper panel of Fig. (\ref{fignew1})), it is evident that the 
form of the \pdf{} 
is still controlled by the saturation mixing ratio profile. But the details, such as the
slope of the \pdf{} for intermediate values of $Q$, are dependent on the eddy-diffusivity and 
forcing. Also, it is 
worth noting that the \pdf{}  
is in marked contrast to what one would obtain for a fully saturated domain. 
For
example in
the saturated case we had $P(q,y)=\delta(q-q_s(y))$, this yields
$\pdf{}(Q_{\textrm{sat}})
\sim - dF(Q)/dQ$ which implies a power law ($Q^{-1}$, with no "roll-off" for high $Q$) when $q_s=\exp(-\alpha y)$. 
At first sight the roll-off in (\ref{11b}) appears to be an obvious result of enforcing rapid 
condensation, but note that the saturated case considered above also conforms with the rapid condensation 
limit yet it yields a very different \pdf{}. This is illustrated more clearly by choosing $q_s(y)$ to be a 
linearly decreasing function of $y$, now the saturated \pdf{} is uniform, whereas 
(\ref{11a}) again yields a slowly decaying \pdf{} with a smooth roll-off at large $q$. \\

With regards to water vapor, given the logarithmic dependence of the infrared cooling on the water vapor mixing ratio,
it follows that fluctuations in the mixing ratio actually increase infrared cooling \cite{Ray-2}. 
Hence a slowly decaying \pdf{}, by increasing the
probablity of encountering a large fluctuation (as compared to a normal \pdf{}), increases the efficiency of radiative cooling to
space. In this regard, the rapid roll-off of the tail of the \pdf{} could play a strong role in that it 
sharply decreases the possibility of very 
large fluctuations --- for example, in comparison to a advection-diffusion model of water vapor distribution where one 
would encounter
exponential tails under homogeneous forcing \cite{Chert} --- 
and would imply a reduction in the infrared
cooling or in other words an increase in the temperature of the surface to maintain energy balance. 
Quantifying this effect by using a full-fledged atmospheric radiation code is a project we hope to pursue in the near future. \\

\section{Numerical Simulations}

A primary assumption in our derivation was the use of an "eddy diffusivity" to represent the effect of 
an advecting velocity field. As mentioned, in most flows of interest the velocity field is expected to be quite 
coherent and might not conform to a diffusive representation. To test the stringency of this assumption
we numerically 
simulate the ADC system by employing a lattice map for 
purposes of smooth large scale advective mixing \cite{Ray},\cite{jai2}. The velocity field is

\begin{eqnarray}
u(x,y,t) = f(t)~A_1 ~ \textrm{sin}(B_1 y + p_n) \quad  \nonumber \\
v(x,y,t) = (1-f(t))~A_2 ~ \textrm{sin}(B_2 x + q_n)  \nonumber \\
\label{1m}
\end{eqnarray}
where $A_1=0.75,A_2=0.5,B_1=B_2=2.5$, $f(t)$ is 1 for $nT \le t < (n+1)T/2$ and 0 for $(n+1)T/2 \le t < (n+1)T$. 
$p_n , q_n$ ($\in [0,2\pi]$) are random numbers selected
at the beginning of each iteration, i.e., for each period $T$. Advection is implemented via sequential 
integer shifts in the $x$ and $y$ directions on a square lattice (see \cite{Ray} for details).
Apart from its numerical efficiency the lattice map has the advantage of
preserving moments, i.e. it does not introduce spurious diffusion into the problem. \\

\begin{figure}
\includegraphics[width=7.5cm,height=9cm]{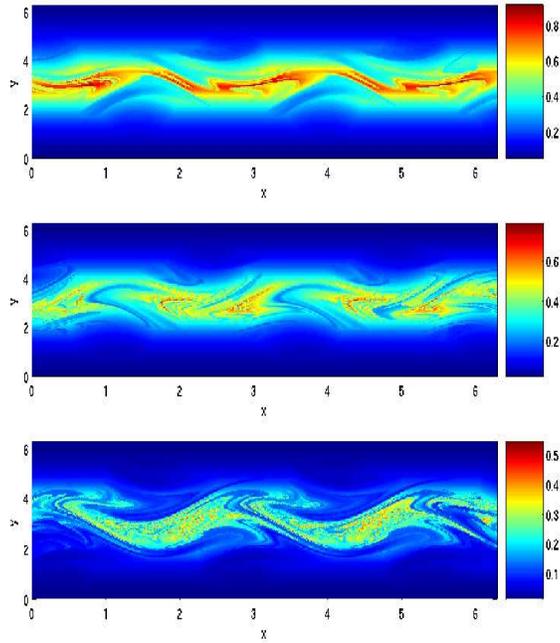}
\caption{\label{fignew22} A snapshot of the condensable field under the action of the 
map specified in (\ref{1m}). 
The snapshots are after the \pdf{} has settled into an invariant shape. The upper,middle and lower panels 
correspond to
$\alpha=1,1.25 $ and $1.5$ respectively with $\delta=\min{(q_s)}$.}
\end{figure}

\begin{figure}
\includegraphics[width=7.5cm,height=7.5cm]{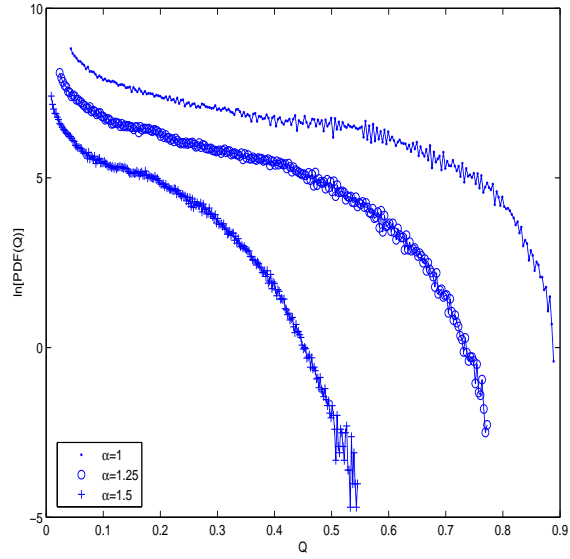}
\caption{\label{fignew3} The \pdf{}'s of the condensable field for $\alpha=1,1.25 $ and $1.5$.
The curves have been shifted for clarity. Note, as per Fig. (\ref{fignew1}), as $\alpha$ increases 
$\delta$ decreases resulting in an increase in the slope 
(as $\kappa_e$ is the same in every case) of the \pdf{}. Also notice the 
increase in the sharpness of the peak with $\alpha$.}
\end{figure}

To test the features of the \pdf{} seen in Fig. (\ref{fignew1}), we set $\delta=\min{(q_s)}/\Delta t$. 
Note that 
as the advective map has no time scale, we set $\Delta t= 1$ to match an iteration of the mapping.
Numerically this amounts to incrementing the mixing ratio at every site on the lattice by 
$\min{(q_s)}$ at every iteration. Now as $\min{(q_s)}$ is a function of $\alpha$, $\delta$ decreases with increasing $\alpha$.
Interestingly, we observe that even though the maximum attainable mixing ratio is
$\max{(q_s)}$, the system achieves equilibrium for much smaller maxima in $q$ field. 
A snapshot of the equilibrium condensable field for $\alpha=1,1.25$ and $1.5$ is shown in 
Fig. (\ref{fignew22}) --- note the increase in "graininess" of the field with $\alpha$. 
The \pdf{}'s for these cases can be seen in 
Fig. (\ref{fignew3}) 
--- the plotted \pdf{}'s
are averages over the last couple of iterations of the map.
In spite of the non-local nature of the mixing protocol the shapes of
the \pdf{}'s display the features anticipated from the eddy-diffusive case, in particular we see the decrease in 
the slope of the \pdf{} with $\delta$ (the advective map is unchanged so $\kappa_e$ is the same in all the simulations)
also the characteristic roll-off induced via the form of the saturation profile in conjunction with the 
limit of rapid condensation is evident. \\

\begin{figure}
\includegraphics[width=7.5cm,height=7.5cm]{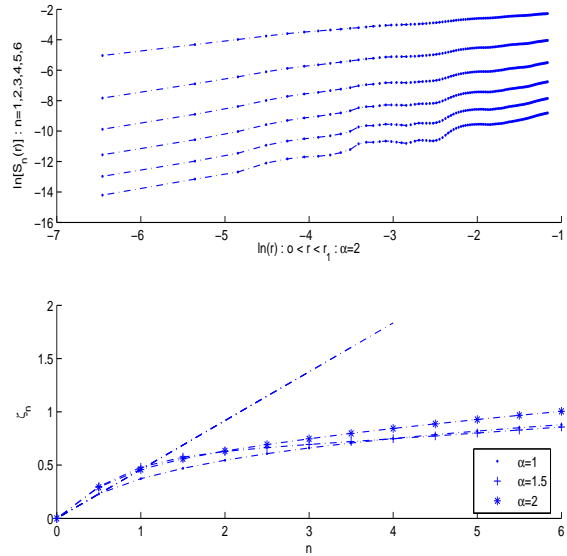}
\caption{\label{fig_s} Upper Panel : Plots of $\log(S_q(r))$ Vs. $\log(r)$ for the first six
moments of the equilibrium condensable field with $\alpha =2$. 
Lower panel : Scaling exponents extracted from above (for $\alpha=2$) and similar plots (not shown) for 
$\alpha=1,1.5$.} 
\end{figure}

Another measure of quantifying intermittency in a field is to examine its structure functions, defined as 
$S_n(|\vec{r}|) = < |q(\vec{x}+\vec{r})-q(\vec{x})|^n >$ (in the present situation $<\cdot>$ denotes a spatial average) 
higher moments (i.e. larger $n$) of the structure functions are most sensitive to 
the "roughest" regions of the 
field. 
For the condensable field, we observe that $S_n(r) \sim r^{\zeta_n}$ for $l_1 \le r \le l_2$ where $l_1 \rightarrow 0$ as 
we are dealing with a non-diffusive problem and $l_2$ is an outer 
scale which is smaller than the 
size of the domain. More to the point, the exponents $\zeta_n$ are anomalous in 
the sense $\zeta_n < n \zeta_1$ for $n>1$. In fact, we observe near saturation of the scaling exponents (i.e. $\zeta_n$ tends to 
a constant) for large $n$. Both the scaling of the structure functions and the extracted scaling exponents are 
shown in Fig. (\ref{fig_s}). 
The anomalous behaviour is physically anticipated as smooth advection by itself leads to the formation of 
sharp fronts in finite time --- hence 
one has a field composed of smooth regions
interrupted by step like discontinuities. Indeed the combination of these two structures yields a $\zeta_n$ profile which
increases linearly for $n<=1$ and then saturates to a constant (i.e. extreme intermittency) for $n>1$ \cite{Aurell}. 
As the additional presence of condensation 
does not involve any
smoothening of the field, we do not expect it to be able to mollify the discontinuities created via advection.
But rapid condensation provides a bound for the magnitude of the jump across the 
advective discontinuity, i.e. $|q(x+r)-q(x)| \le 1-\exp(-\alpha L/2)$, hence we expect a weak dependence of 
$\zeta_n$ on $\alpha$. 
Indeed, this qualitative reasoning is bourne out in the lower panel Fig. (\ref{fig_s}).
In a similar vein, it has been shown 
(under further assumptions regarding the nature of the velocity field) 
that an interplay of advection with linear damping produces severe anomalous scaling \cite{Chertkov1}.
Interestingly, such anomalous behaviour --- with saturation of $\zeta_n$ for large $n$ --- 
has been observed in an analysis of 
specific humidity fluctuations from high resolution aircraft data in the troposphere \cite{cho}. \\


\section{Conclusions and Atmospheric Data}

We have examined the state of equilibrium 
achieved via the interplay of smooth advection and condensation.
Exploiting the attainment of equilibrium without the presence of gradient fields allows
us to make progress on the equation governing the \pdf{} of the condensable substance. 
In particular, the limit of rapid condensation admits a simplified Liouville equation 
with an integral normalization constraint on the \pdf{}. Assuming a straightforward 
uniform solution to the governing equation shows the \pdf{} to be tied to the particular
saturation profile under consideration. Relaxing this uniformity and assuming an 
eddy-diffusive nature for the advective velocity fields shows the form of the \pdf{} to still
be controlled by the saturation profile but details, such as the rate of decay, to be 
sensetive to the eddy-diffusivity and strength of the forcing. Indeed, numerical simulations employing a 
lattice map for advective purposes reproduce these features.
Further, scaling exponents extracted from 
structure functions of the condensable field show anomalous behaviour. Physically, 
the anomalous scaling is 
anticipated given the tendency of undiffused smooth
advection to create sharp fronts. In fact, the near saturation of $\zeta_n$ for large $n$ is a reflection of these
fronts providing the leading contribution to the higher order structure functions. \\

\begin{figure}
\includegraphics[width=7.5cm,height=7.5cm]{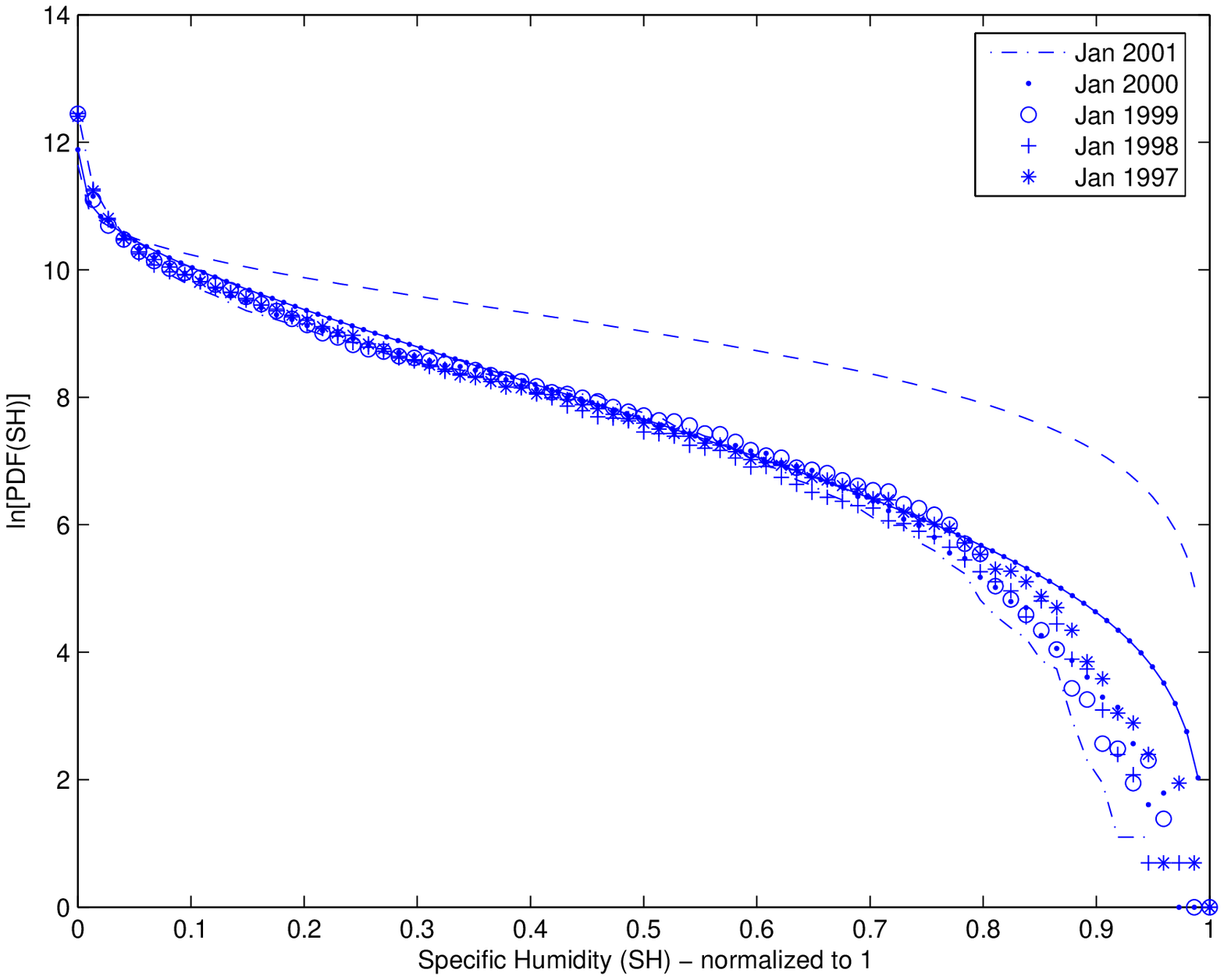}
\caption{\label{fig4} The normalized specific humidity \pdf{}'s in the midlatitudes (between $30^{\circ}$
and $60^{\circ}$ in both hemispheres) along the 
300 K isentropic surface
for Jan 97-01 from ECMWF data. The solid line with dots is a plot of (\ref{11b}) (shifted vertically by a constant) 
with $\frac{\kappa_e}{\delta} = 45,
\alpha =3$ and for comparison the dash-dash line is a plot of the \pdf{} resulting from uniformity, again with $\alpha=3$.}
\end{figure}

As mentioned in the Introduction, our motivation is to understand the distribution of water vapor 
along midlatitude isentropic surfaces. With this in mind we construct \pdf{}'s of the 
midlatitudinal specific humidity
field along the 300 K
isentrope using data from the ECMWF reanalysis (ERA40) project. As is seen in Fig. (\ref{fig4}),
despite the simplifying assumptions made in our derivation,
the \pdf{}'s from data can be approximated by the eddy-diffusive \pdf{}. In fact, in
Fig. (\ref{fig4}) we've also plotted the \pdf{} that results from assuming uniformity (for the same 
$\alpha$)
and as is seen even though the primary shape of the data \pdf{} follows from the saturation profile in
conjunction with rapid condensation, it is the enhanced slope via $\frac{\kappa_e}{\delta}$ that captures
the decay of the \pdf{}.
Note that for large $q$ the \pdf{}'s from 
data deviate significantly from the eddy-diffusive estimate. 
As we have considered the nondiffusive limit a possible 
source of this discrepancy might lie in the homogenization induced by diffusion. Secondly, as 
the source of water vapor lies in the tropics a boundary forcing might be more appropriate for
the actual atmospheric problem.
Even so, these moderately encouraging results lead us to conjecture that
the ADC model --- in the limit of rapid condensation with the proper saturation profile --- driven by idealized velocity fields
might be of use in predicting the statistical properties of the large scale distribution of water vapor
in the midlatitude troposphere. \\

\acknowledgments

We thank Prof. W.R. Young (Scripps Institute, UCSD) for his suggestions 
which led to a clearer formulation of the problem.
J.S. would also like to acknowledge helpful conversations with Dr. A. Alexakis (NCAR). The comments of the
anonymous referee are gratefully acknowledged, in particular they distinctly improved the formulation of the \pdf{} equation. We also thank the referee for pointing out the resemblance to the
spiking neuron problem.
Much of this work was carried out while the first author was at the National Center for Atmospheric Research
which is sponsored by the National Science Foundation.  The second author's contribution
to this work was sponsored by the National Science Foundation under grant ATM-0123999.

\clearpage

\begin{thebibliography}{99}

\bibitem{Falk} G. Falkovich, K. Gawedzki and M. Vergassola, 
Rev. Mod. Physics, {\bf 73}, 913 (2001).

\bibitem{Sherwood} S.C. Sherwood,
J. Climate, {\bf 9}, 2919 (1996).

\bibitem{SH} E.P. Salathe and D.L. Hartmann,
J. Climate, {\bf 10}, 2533 (1997).

\bibitem{Ray-grl} R.T. Pierrehumbert, Geophys. Res. Lett. {\bf 25}, 151 (1998).

\bibitem{Ray-Nature} R.T. Pierrehumbert, Nature, {\bf 419}, 191 (2002).

\bibitem{HS} I. Held and B. Soden, 
Annu. Rev. Energy Environ. {\bf 25}, 441 (2000).

\bibitem{Ray-2} R.T. Pierrehumbert, H. Brogniez and R. Roca; 
in {\it The General Circulation of the Atmosphere}, edited by T. Schneider and A. Sobel (Princeton
University Press, 2005) (to appear)

\bibitem{Chertkov1} M. Chertkov, Phys. of Fluids,
{\bf 10}, 3017 (1998).

\bibitem{Neufeld} Z. Neufeld, C. Lopez and P.H. Haynes, Phys. Rev. Lett. {\bf 82}, 2606 (1999).

\bibitem{Abr} E.R. Abraham, Nature, {\bf 391}, 577 (1998).

\bibitem{Bert} L. Berthier, J-L. Barrat and J. Kurchan, 
Phys. Rev. Lett. {\bf 86}, 2014 (2001).

\bibitem{Hoskins} B. Hoskins, 
Tellus, {\bf 43A}, 27 (1991).

\bibitem{Ray-4} H. Yang and R.T. Pierrehumbert, 
J. Atmos. Sci. {\bf 51}, 3437 (1994).

\bibitem{Cycle} P. Cvitanovi\'c, R. Artuso, R. Mainieri, G. Tanner and G. Vattay,
{\em Chaos: Classical and Quantum},
{\tt ChaosBook.org} Niels Bohr Institute, Copenhagen (2005).

\bibitem{Fusi} S. Fusi and M. Mattia,
Neural Computation {\bf 11}, 633 (1999).

\bibitem{P} S.B. Pope, Prog. Energy Combust. Sci. {\bf 11}, 119 (1985). 

\bibitem{Dopazo-rev} C. Dopazo, L. Valino and 
N. Fueyo, Int. Journal of Mod. Physics B, {\bf 11}, 2975 (1997); 

\bibitem{jai1} J. Sukhatme, Phys. Rev. E,
{\bf 69}, 056302 (2004).


\bibitem{MK} A.J. Majda and P.R. Kramer, Physics Reports, {\bf 314}, 238 (1999). 

\bibitem{jai3} J. Sukhatme, J. Atmos. Sci. {\bf 62}, 3831 (2005).


\bibitem{Chert} M. Chertkov, G. Falkovich, I. Kolokolov and V. Lebedev, Phys. Rev. E, {\bf 51}, 5609 (1995).

\bibitem{Ray} R.T. Pierrehumbert, 
Chaos, {\bf 10}, 1, 61 (2000). 

\bibitem{jai2} J. Sukhatme and R.T. Pierrehumbert, Phys. Rev. E, {\bf 66}, 056302 (2002).

\bibitem{Aurell} E. Aurell E, U. Frisch, J. Lusko and M. Vergassola, J. Fluid. Mech., {\bf 238}, 467 (1992).

\bibitem{cho} J. Cho, R. Newell and G. Sachse, Geophys. Res. Lett. {\bf 27}, 377 (2000).


\end{thebibliography}
\end{document}